\documentclass{elsart}

\usepackage{amssymb}
\usepackage{epsfig}
\usepackage{natbib}

\begin{document}

\begin{frontmatter}



\title{X-ray Timing beyond the Rossi X-ray Timing Explorer}


\author{Didier Barret}

\address{Centre d'Etude Spatiale des Rayonnements, 9 Avenue du Colonel Roche, 31028 Toulouse Cedex 04, France}

\begin{abstract}
With its ability to look at bright galactic X-ray sources with sub-millisecond time resolution, the Rossi X-ray Timing Explorer (RXTE) discovered that the X-ray emission from accreting compact stars shows quasi-periodic oscillations on the dynamical timescales of the strong field region. RXTE showed also that waveform fitting of the oscillations resulting from hot spots at the surface of rapidly rotating neutron stars constrain their masses and radii. These two breakthroughs suddenly opened up a new window on fundamental physics, by providing new insights on strong gravity and dense matter. Building upon the RXTE legacy, in the Cosmic Vision exercise, testing General Relativity in the strong field limit and constraining the equation of state of dense matter were recognized recently as key goals to be pursued in the ESA science program for the years 2015-2025. This in turn identified the need for a large (10 m$^2$ class) aperture X-ray observatory. In recognition of this need, the XEUS mission concept which has evolved into a single launch L2 formation flying mission will have a fast timing instrument in the focal plane. In this paper I will outline the unique science that will be addressed with fast X-ray timing on XEUS.\end{abstract}

\end{frontmatter}

\section{Fast time variability and strong gravity}
\begin{figure}
\centerline{\psfig{figure=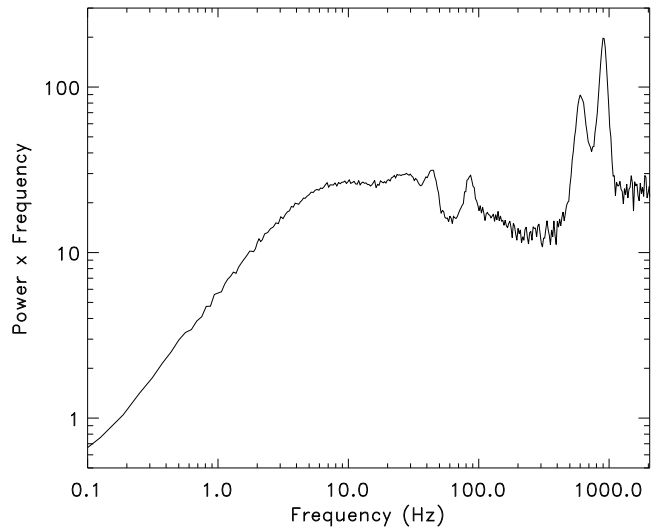,height=3.0in}}
\caption{Power density spectrum of Sco X-1 in a $\nu F\nu$-like representation, showing two kHz quasi-periodic oscillations (data courtesy of Michiel van der Klis).
\label{fig:sco}}
\end{figure}

The X-ray emission from accreting compact stars (neutron stars and black holes) has been shown to vary on (sub)-millisecond timescales, i.e. on timescales comparable to the dynamical timescales of the innermost regions of the accretion disk. This variability was discovered with RXTE \citep{bra93} just ten years ago in the form of kilo-Hz quasi-periodic oscillations (QPOs) in Fourier power density spectra (see the review by van der Klis (2005) and Fig. \ref{fig:sco}). There is a wide consensus that these QPOs probe the motion of matter under strong gravity with the signal originating from within a few Schwarzschild radii of the compact star. This is a region where the spacetime curvature is extreme, orders of magnitude larger than that sampled in weak field tests (e.g. Gravity-Probe B). This is a regime where the most dramatic effects of GR are expected, and where fundamental predictions of GR, such as the existence of an innermost stable circular orbit (ISCO) have yet be tested. With RXTE, X-ray timing has thus become a complementary tool to X-ray spectroscopy, polarimetry and gravitational wave measurements to probe strong field GR. It has even been claimed that studying black hole variability might be used to test alternative theories of gravity \citep{psa04}.
\subsection{Overall QPO properties}
It is is beyond the scope of this paper to review the properties of kHz QPOs, and the reader is referred to van der Klis (2005) for a recent review and {\it X-ray Timing 2003 - Rossi and Beyond} by Kaaret, Lamb and Swank (2004, AIP Conference Proceedings, Vol. 714, Melville, NY: American Institute of Physics) on which the present paper draws heavily on. Here I will focus on their general properties, relevant to some of the points discussed below. So far, kHz QPOs have been detected from about 20 neutron star low-mass X-ray binaries (about 10\% of known systems) \citep{swa04}, and from a handful of binaries containing a black hole candidate.  In general two kHz QPOs are observed. The properties of the QPO (frequency, amplitude, width) vary in a complicated way with parameters such as the source luminosity. For neutron stars, the QPO frequencies can vary by at least two orders of magnitude with the amplitude decreasing with increasing frequency. There is a trend for the QPO fractional amplitude to increase with energy, reaching in some cases 20\% at $\sim 20$ keV. In black holes, QPOs are detected closer to the sensitivity limit of RXTE with typical amplitudes of 1\%: they are transient, show stable frequencies and are detected in the hardest X-ray bands (typically above 10 keV)  \citep{rem04}. 

\subsection{QPO interpretation}
The true nature of the underlying signal is unknown and there is a wide range of possible models to be considered, mostly focussing on predicting QPO frequencies. Because the disk is a natural source of periodicities, from orbital to epicyclic motions and disk oscillations, it is generally agreed that QPOs originate in the cool ($\sim 1-2$ keV) accretion disk. This implicitly assumes that the oscillator must survive the strong damping forces thought to be present in the disk and that an amplification mechanism operates to account for the fact that QPOs show larger amplitude at higher energies ($\sim 10-20$ keV). For neutron star QPOs, the main constraint for the models so far comes the observation that the frequency separation between the twin QPOs is close to the spin frequency or half the spin frequency of the neutron star (where that is known). This strongly suggests that the spin is involved in the generation of the QPOs. For black holes, the main constraint for the models comes from the observations that QPOs, when observed in pairs, have commensurate frequencies, with ratios close to small integer ratios (e.g. 3:2) \citep{rem04}, lending support to the idea that a resonance mechanism is involved in the make up of the QPOs. 

Some models (not all of them consistent with the constraints above) relate QPO frequencies directly characteristic frequencies of test particles moving in the strongly curved spacetime (General Relativistic orbital and epicyclic frequencies \citep{abr04}),  others link the QPO frequencies to the frequency of orbiting clumps interacting with radiation from the neutron star \citep{mil98}. Some models involve global disk oscillations, relativistic \citep{kat01}, or not \citep{tit04}. Yet others associate QPOs with strong field GR effects, e.g. relativistic dragging of inertial frames \citep{ste99}, see \citep{kli05} for a more complete list of references. 

There is however a trend in the models to associate QPOs with resonance phenomena, involving General Relativistic frequencies. As an example, in the accreting millisecond pulsar, SAXJ1808-3658, it has been proposed that the two QPOs (whose frequency difference is just equal to half the neutron star spin frequency) are generated at a radius in the disk where the difference between the general relativistic orbital frequency and radial epicyclic frequency is equal to half the spin frequency \citep{wij03}. Similarly, to account for the 3:2 frequency ratios observed in black hole systems, a parametric resonance concept has been put forward, in which the QPOs are produced at a radius in the disk where two of the three general relativistic frequencies (orbital, vertical and radial epicyclic) have commensurate values, matching the observed QPO frequencies \citep{abr04}. Although this is still very much under discussion, it is interesting to note that the latter two models link QPOs directly to general relativistic frequencies.
\subsection{The innermost stable circular orbit}
The ISCO is one of the key predictions of strong gravity GR stating that there exists a region around sufficiently compact stars within which no stable circular orbital motion is possible. In a Schwarzschild geometry, the radius of the ISCO is 6GM/c$^2$, which is larger than the radii of
   neutron stars, deduced from models constructed with most modern high-density equations 
   of state \citep{akm98}. Interestingly enough, claims have already been made that the QPO properties have revealed signatures of the ISCO. Prior to the launch of RXTE, it was suggested that the ISCO might induce a frequency cutoff in the power density spectrum \citep{klu90}.  After the discovery of kHz QPOs, it was proposed that signatures of the ISCO could include a frequency saturation of kHz QPOs with increasing mass accretion rate, or a drop in the quality factor and amplitude of kHz QPOs as they approach a limiting frequency \citep{mil98}. A saturation of the QPO frequency with count rate \citep{zha98} or inferred mass accretion rate \citep{blo00}, as well as a sudden drop of coherence and amplitude at some critical frequency \citep{bar05b} have now been reported. Although the interpretation of these results is still debated, this demonstrates the potential of fast X-ray timing for testing the existence of the ISCO. Note that identifying securely a frequency with an orbital frequency at the ISCO would yield directly the mass of the neutron star.
\subsection{QPO and the black hole spin and mass}
\begin{figure}
\centerline{\psfig{figure=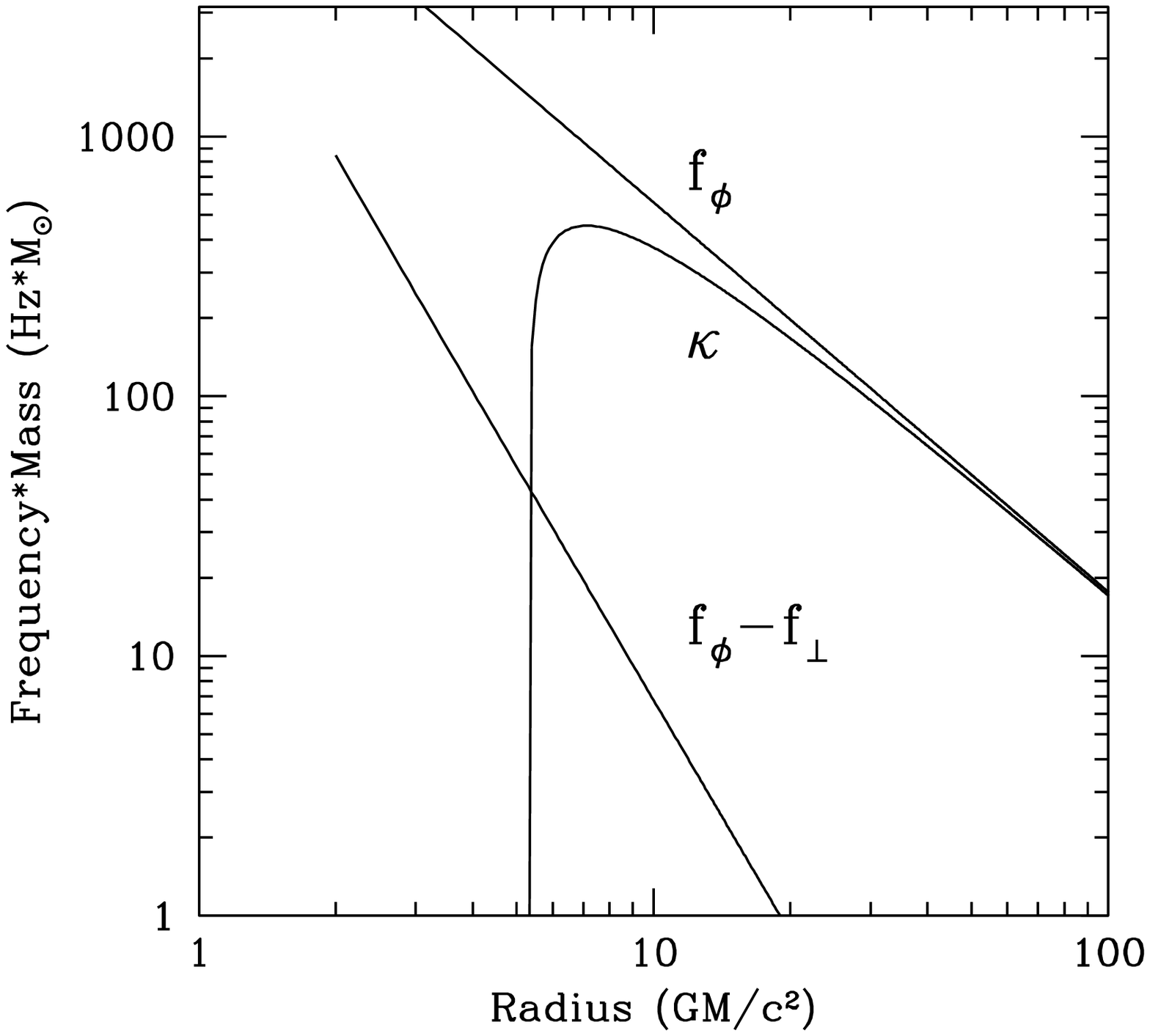,height=2.5in}\psfig{figure=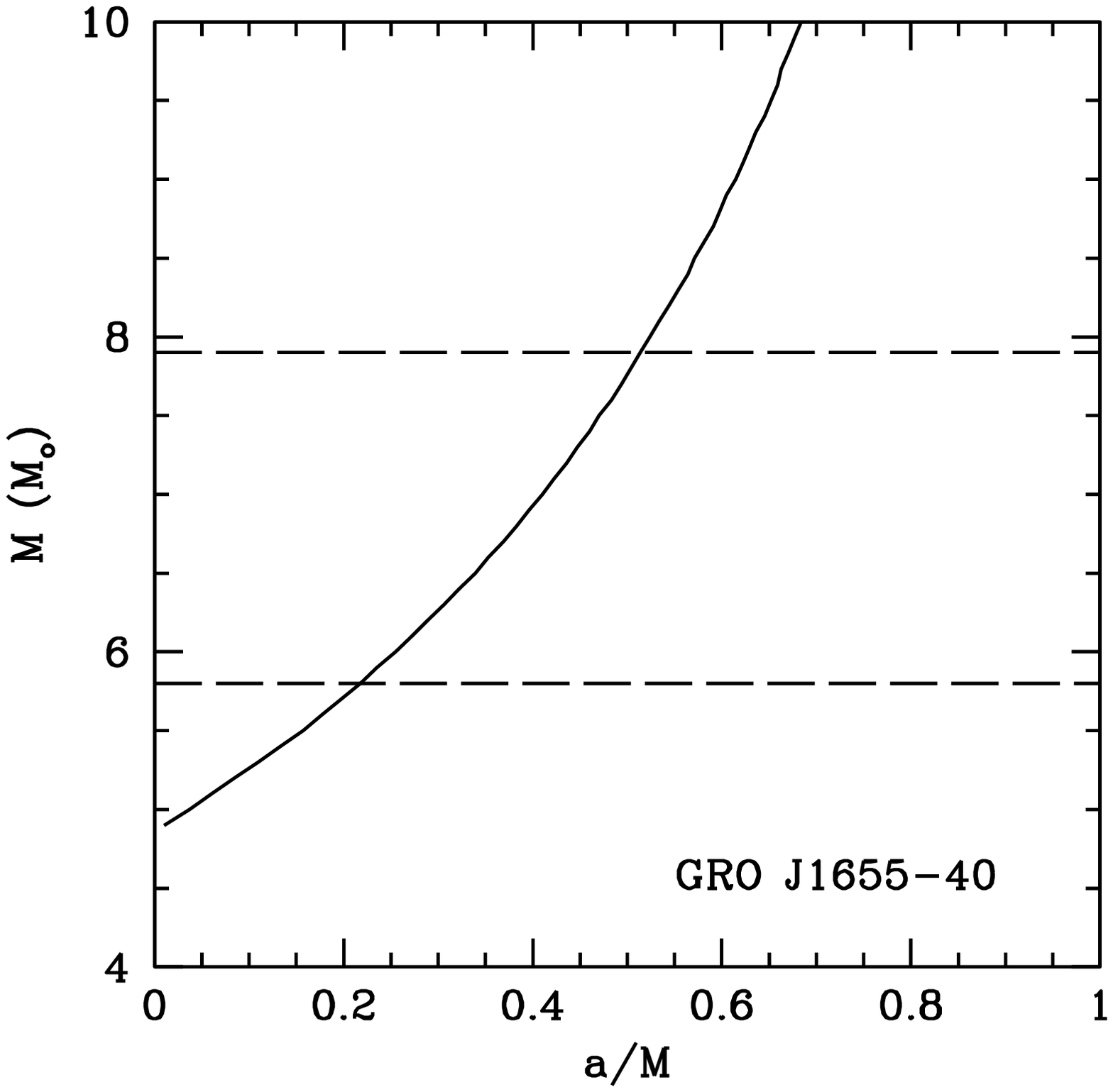,height=2.5in}}
\caption{Left) The azimuthal ($f_\phi$), radial epicyclic ($\kappa$), 
and vertical ($f_\perp$) frequencies at different radii around a
black hole with spin parameter $a/M=0.2$. Right) {\em Dashed Lines:\/} the dynamical measurement of the black-hole mass in the source GRO~1655$-$40. {\em Solid Line:\/} the
minimum spin parameter $a/M$ for each black-hole mass, for which the observed 450~Hz QPO can be produced as a Keplerian frequency of a stable orbit.  According to this argument, the black hole in GRO~1655$-$40 is spinning with $a/M>0.2$ (Both figures from \citep{psa04})
\label{fig:psa}}
\end{figure}

Depending on the interpretation, parameters such as the spin and the mass of black holes can be constrained directly from measuring the QPO frequencies. One example is given in Fig. \ref{fig:psa} where the most model independent constraint is used. It simply assumes that the QPO frequency is smaller than the orbital frequency at the ISCO (the latter being a function of the black hole mass and spin). More stringent constraints on the black hole mass and spin can be obtained from models involving resonance between relativistic frequencies \citep{psa04,abr04}. All constraints point to maximally spinning black holes. As an application, because of the 1/M dependency of all general relativistic frequencies, it has been proposed that measuring QPOs from Ultra-Luminous X-ray sources, whose nature is a mystery, might help in constraining the mass of the compact object \citep{abr04} in these systems.
\section{Fast X-ray timing and matter at supra-nuclear density}
Besides providing a laboratory to test GR,  neutron stars have also other interesting properties; 
some may be powerful emitters of gravitational waves at their millisecond spin periods, some have 
an extreme magnetic field in which strong-field QED  effects can be studied. But the overall reason why 
neutron stars have stimulated much work is because the physical state of the matter which makes up their cores is still a mystery. The density of a neutron star can reach up to 10 times the density of a nucleus; besides neutrons and protons, at such high densities the physics of strong interactions predicts the existence of hyperons - particles that contain strange quarks - pions and kaon condensates, or even free quarks. Neutron stars offer the only possibility of detecting such exotic matter in the bulk, and resolving the question of the most stable form of matter (baryons or free quarks). Since the structure of neutron stars is set by the, yet unknown, equation of state of matter at these high densities, measurements of neutron stars masses and radii can provide information not only on the physical makeup of neutron stars, but also on the nature of the interactions between the particles that constitute them. The equation of state of nuclear matter is at the heart of many fundamental problems in physics, from nuclear physics to condensed matter physics, including superconductivity and super-fluidity. Fast X-ray timing and high resolution spectroscopy (e.g. redshifted lines emitted at the neutron star surface) holds great potential for constraining the physics of strong interactions.

\subsection{QPOs and the equation of state of dense matter}
\begin{figure}
\centerline{\psfig{figure=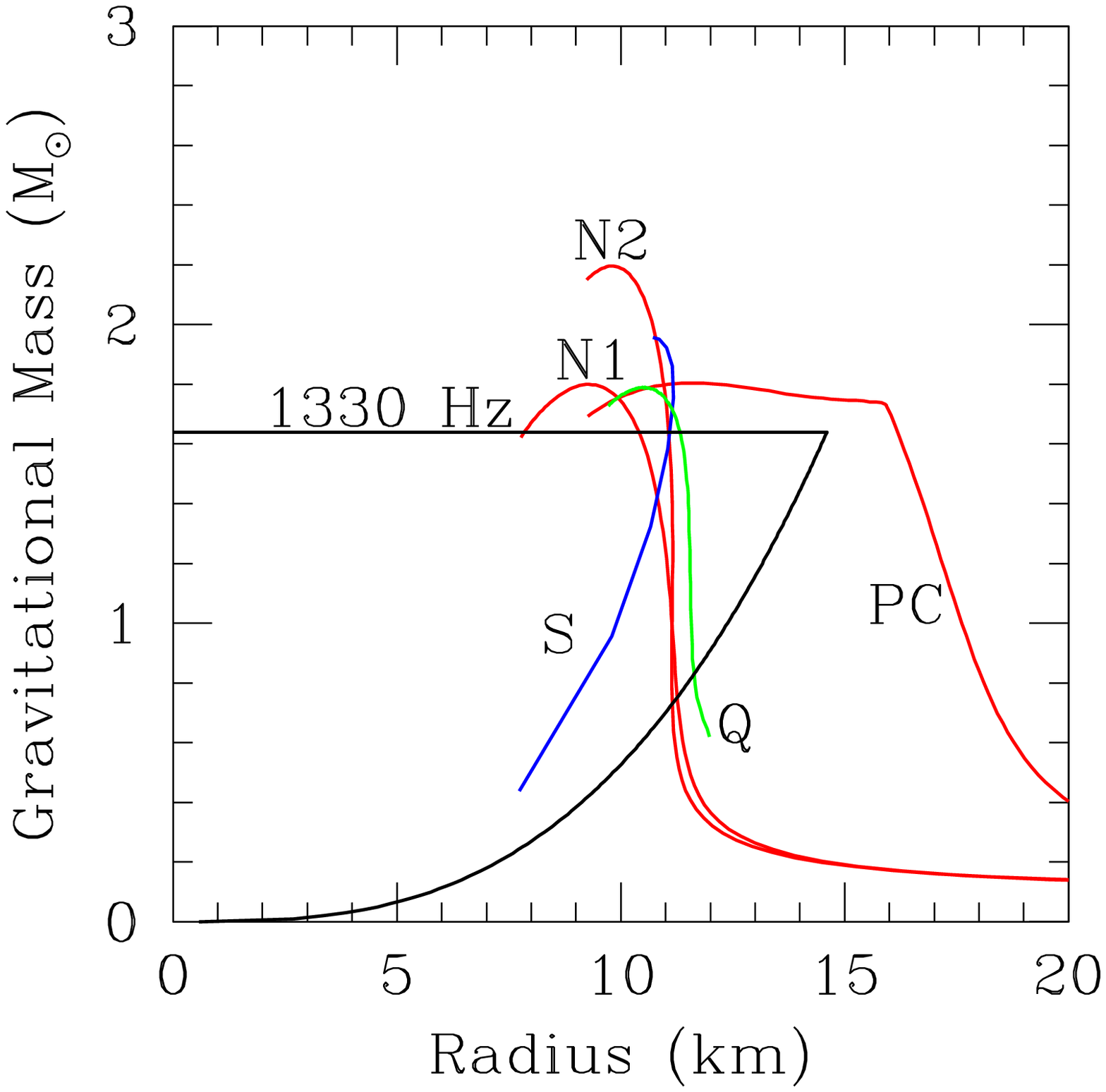,height=3.0in}\psfig{figure=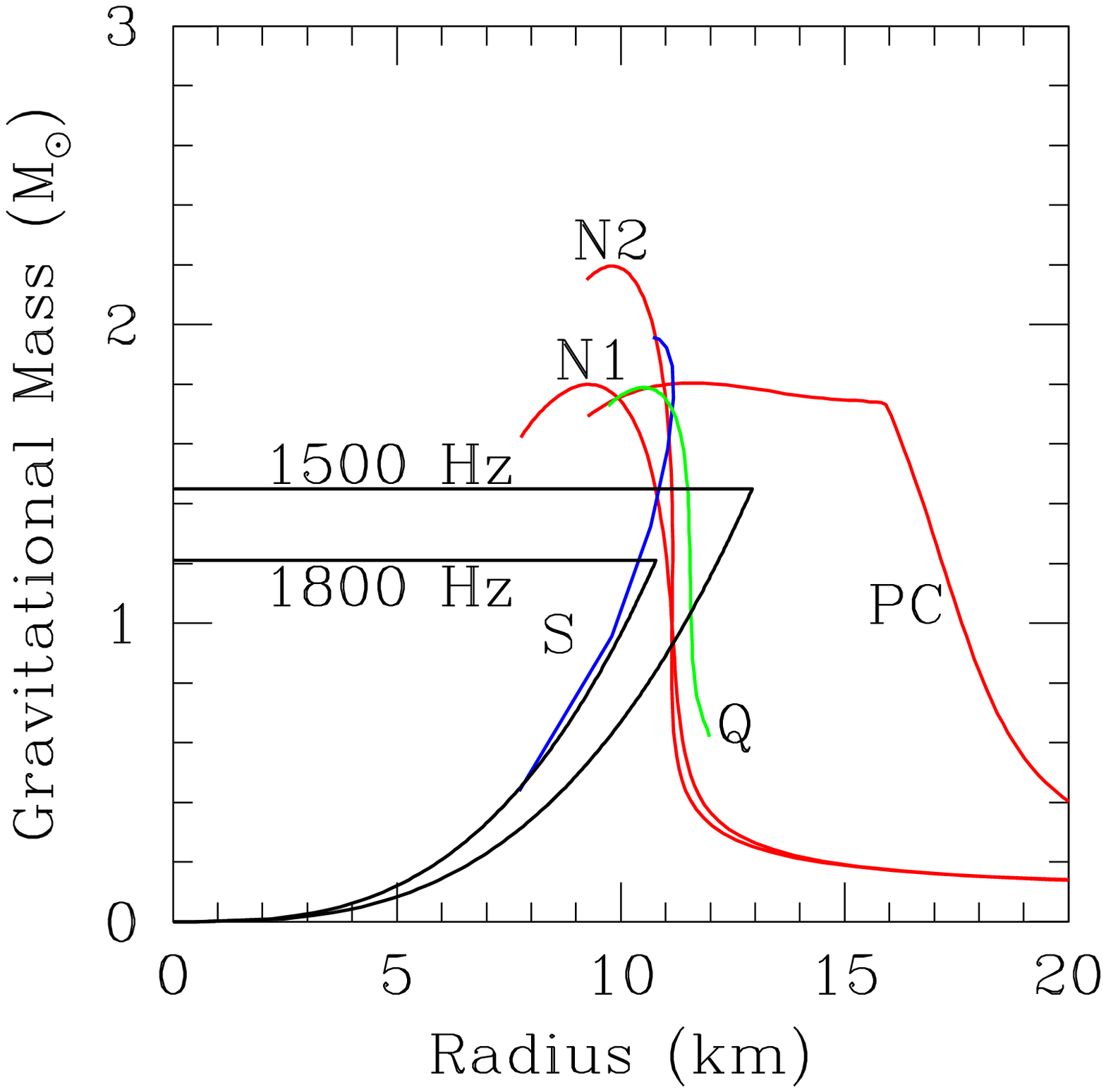,height=3.0in}}
\caption{left) Constraints from orbital frequencies.  The 1330~Hz curve is for
the highest kilohertz quasi-periodic oscillation frequency yet measured.  This curve is for a
non-rotating star; the constraint wedge would be enlarged slightly for a
rotating star.  The solid lines are mass-radius curves for different representative high-density equations of
state. Right) Constraints on mass and radius for hypothetical detections of a 1500~Hz QPO and a 1800~Hz QPO, if they are identified with an
orbital frequency.  At 1500~Hz one expects signatures of the ISCO to
be present; a detection of 1800~Hz would present strong difficulties
for standard nucleonic equations of state.  Both figures taken from Miller (2004). \label{fig:mil}}
\end{figure}

If QPOs arise from stable orbital motion (the highest frequency QPO is generally associated with an orbital frequency in most models), the frequency yields the orbital radius at which it is produced. Hence this orbital radius limits the mass and the radius of the neutron star \citep{mil98}. This is because two types of constraints can be set. The first one uses the fact that no QPOs are expected to be produced within the ISCO, hence the orbital radius must be larger than the radius of the ISCO (6GM/c$^2$ in a Schwarzschild geometry). The second constraint simply states that the radius of the neutron star is smaller than the orbital radius. So far, the highest QPO frequency is 1330 Hz, limiting the mass and radius of the neutron star in the bottom left area of Fig \ref{fig:mil}. 
\subsection{Waveform fitting of neutron star X-ray oscillations and dense matter}
The discovery of the first accreting millisecond pulsar (a handful are known today) and transient oscillations in type I X-ray bursts in a dozen sources have provided us with additional tools to constrain dense matter. This is because these pulsations are likely caused by hot spots rotating at the surface of the neutron star \citep{mun04,str04}. The emission from the spot will travel in a strongly curved space time before reaching the observer, and will be subject to a variety of effects, including relativistic aberration, Doppler boosting and strong field gravitational light bending \citep{pou04,str04}. The more compact the star, the smoother will be the pulse profile. Fitting of the pulse profile of the oscillations with generic light curves computed under specific assumptions (spacetime metric, geometry, emission pattern and spectrum of the spot,....) has already demonstrated the potential of this approach to constrain the compactness of the star, especially when the harmonic content of the oscillations is also detected \citep{str04,bat05}.

\section{From RXTE to XEUS: From 0.7 m$^2$ to $\sim 10$ m$^2$}
As stressed above, we are at the beginning of a new era, which means that the full potential of the RXTE discoveries is difficult to anticipate. It is however clear that fast X-ray timing holds promises to address some of the most important issues of modern astrophysics, such as strong gravity, accretion disk physics, and  the physics of the strong interactions. The obvious path forward will therefore require an instrument with a significant increase in sensitivity compared to RXTE. For timing, the significance of a QPO detection scales proportionally with the number of photons collected per unit time, hence the with collecting area of the instrumentation. Given that RXTE has a collecting area of $\sim 0.7$ m$^2$, a $\sim 10$ m$^2$ class X-ray telescope (with say 3 m$^2$ effective area at 6 keV) has the potential to provide an improvement in sensitivity for timing measurements by more than one order of magnitude over RXTE. Such an instrumentation would allow breakthrough science in several ways. The most important ones are listed below.
\subsection{Breakthrough}
It is already clear that with RXTE, we have only seen the tip of the iceberg of the high frequency phenomena around compact stars. A factor of 3 to 4 weaker signals in terms of RMS amplitude will be detected with a 10 m$^2$ instrumentation. Side-band frequencies, predicted in some models, will be detected, whose inter-dependence will help in revealing their nature. Measuring simultaneously three frequencies could establish their identification with GR frequencies, if for instance they vary in the way predicted. New analysis techniques, such as the one based on the bicoherence, which measures the degree of phase coupling between oscillations at different frequencies, might be sensitive enough to discriminate between models \citep{mac05}. Furthermore, because the RMS amplitude of the QPO decreases with increasing frequencies, QPOs will be detected at frequencies higher than currently achieved with RXTE. This would allow a search for the frequency ceiling expected at the ISCO and, provided that the orbital nature of the signal is demonstrated, would restrict critically the region of acceptance for neutron star masses and radii, hence allowing some proposed equations of state to be eliminated (see Fig 3).

A second breakthrough would be the determination of the underlying modulation mechanism of the signal. In general, it is simply assumed that a QPO signal is made of sine-wave damped exponential shots, with a time constant (also called coherence time) which relates to the measured width of the QPO in the Fourier spectrum. Coherence time of the order of 0.1 second have been inferred this way \citep{bar05a,bar05b}. Because this estimate is computed by averaging the signal over thousands of cycles it should be considered as a lower limit (frequency drifts will artificially broaden the signal, thus reducing the inferred coherence time). A 10 m$^2$ class instrumentation will enable to study kilo-Hz QPOs on timescales comparable to their coherence times. This will allow one to test whether the signal is persistent or transient, intrinsically aperiodic or made of a superposition of more coherent pulses. The true lifetime of the underlying oscillator may well be one of the most stringent constraints to the models \citep{bar05b}. Depending on the nature of the underlying signal, time-resolved spectroscopy over its coherence time may become possible, yielding information on the radius where the signal originates (e.g. by detecting Doppler shifts). 

The broad iron line seen from stellar mass black holes \citep{min04} provides complementary diagnostics of the strong field region. The profile of the line is also broadened and distorted by relativistic effects (Doppler shift and beaming, gravitational redshift,...). Combining fast timing and moderate resolution spectroscopy will help in testing models and in determining the true nature of the underlying variability. As an example, proving the orbital nature of the signal could be achieved by measuring independently the frequency, from timing, and the radius at which it is generated from continuum or line spectroscopy \citep{kli04,kli05}.

\begin{figure}
\centerline{\psfig{figure=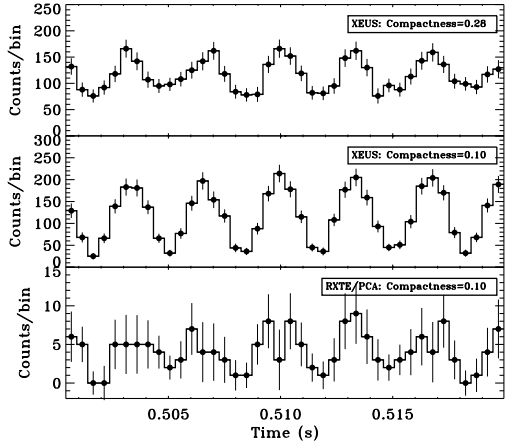,height=2.5in}}
\caption{Simulated light curve of a burst oscillation at 300 Hz as seen by the RXTE/PCA and XEUS. The burst produces 15000 counts/s in the PCA, and 400000 count/s in XEUS. The simulations take into account gravitational light bending in a  Schwarzschild spacetime (one spot and a cosine emission diagram). For XEUS, the light curve is computed for two neutron star compactness (M/R=0.1 and 0.28): the oscillations are directly visible in the light curve even at the highest compactness, whereas they are invisible in the PCA data (Barret 2004). \label{fig:osc}}
\end{figure}

For waveform fitting of X-ray burst and persistent oscillations, a 10 m$^2$ instrumentation would enable the signal to be detected over one cycle (it will also be very sensitive to the harmonic content of the signal). By fitting the waveform, it will be possible to investigate the spacetime around the neutron star, and simultaneously constrain its mass and radius, and hence constrain the equation of state of its high density core with unprecedented accuracy \citep{str04} (see Fig \ref{fig:osc}). 

\subsection{XEUS lining up in the ESA Cosmic Vision}

The ESA Cosmic Vision exercise has defined key goals to be addressed in the science program for the time period 2015 to 2025. Among those goals, the study of matter under extreme conditions (gravity in the strong field limit and the equation of state of dense matter) requires a large aperture X-ray observatory. In parallel, the XEUS mission concept has recently evolved, thanks to the prospects of using light weight Si HPO optics, capable of reaching a 5 arcsec spatial resolution \citep{par03}. Rapid progress is being made and continued development will ensure that a  mirror area of 10 m$^2$ can be flown with a single  launch to the second Earth-Sun Lagrangian point (L2) and autonomous deployment \citep{bav03}. To address specifically the {\it Matter under extreme condition} goal of the Cosmic Vision program, a dedicated timing instrument, capable of coping with the count rates expected from the bright sources of the sky is now in the baseline focal plane instrumentation. The detector implementation described in Barret (2004) is designed to provide an order of magnitude improvement for timing studies over RXTE, while providing more than a factor of 5 better spectral resolution, making possible to combine timing and spectroscopic measurements.
\section{Conclusions}
RXTE has opened a brand new window on fundamental physics. Gravity in the strong field limit and matter at supra-nuclear density, which constitute the core of the RXTE legacy, have been identified as key science goals to be pursued in the science program of ESA after 2015. XEUS, which is foreseen as a potential mission for this time period, now has a dedicated timing detector in the focal plane baseline, with capabilities matching the requirements for a follow-up instrumentation to RXTE. XEUS has therefore a great potential for delivering unique science on strong gravity and dense matter, building on what has been learned in the last few years with RXTE. 
\section*{Acknowledgments}
I wish to thank all colleagues who gave me their support for bringing a fast timing capability in the XEUS focal plane instrumentation. A special thanks to Michiel van der Klis and Mariano Mendez. I am also grateful to Jean-Francois Olive and Gerry Skinner for a careful reading of the manuscript.


\begin{thebibliography}{00}
\bibitem[Abramowicz and Klu{\'z}niak(2004)]{abr04} 
Abramowicz, M.~A., Klu{\'z}niak, W.\ 2004.\ Interpreting black hole QPOs.\ 
AIP Conf.~Proc.~714: X-ray Timing 2003: Rossi and Beyond 714, 21-28. 
 
\bibitem[Akmal et al.(1998)]{akm98} Akmal, A., Pandharipande, 
V.~R., Ravenhall, D.~G.\ 1998.\ Equation of state of nucleon matter and 
neutron star structure.\ Physical Review C 58, 1804-1828. 
 
\bibitem[Barret et al.(2005a)]{bar05a} Barret, D., Olive, 
J.-F., Miller, M.~C.\ 2005.\ An abrupt drop in the coherence of the lower 
kHz quasi-periodic oscillations in 4U 1636-536.\ Monthly Notices of the 
Royal Astronomical Society 361, 855-860. 
 
\bibitem[Barret et al.(2005b)]{bar05b} Barret, D., 
Klu{\'z}niak, W., Olive, J.~F., Paltani, S., Skinner, G.~K.\ 2005.\ On the 
high coherence of kHz quasi-periodic oscillations.\ Monthly Notices of the 
Royal Astronomical Society 357, 1288-1294. 
 

\bibitem[Barret(2004)]{bar04} Barret, D.\ 2004.\ XTRA: The 
fast X-ray timing detector on XEUS.\ AIP Conf.~Proc.~714: X-ray Timing 
2003: Rossi and Beyond 714, 405-412. 

\bibitem[Bavdaz et al.(2003)]{bav03} Bavdaz, M., Peacock, 
A.~J., van der Laan, T., Parmar, A.~N.\ 2003.\ XEUS: approaches to mission 
design.\ X-Ray and Gamma-Ray Telescopes and Instruments for Astronomy.~ 
Edited by Joachim E.~Truemper, Harvey D.~Tananbaum.~ Proceedings of the 
SPIE, Volume 4851, pp.~396-404 (2003).\ 4851, 396-404. 

\bibitem[Bhattacharyya et al.(2005)]{bat05} Bhattacharyya, 
S., Strohmayer, T.~E., Miller, M.~C., Markwardt, C.~B.\ 2005.\ Constraints 
on Neutron Star Parameters from Burst Oscillation Light Curves of the 
Accreting Millisecond Pulsar XTE J1814-338.\ Astrophysical Journal 619, 
483-491. 
 
\bibitem[Bloser et al.(2000)]{blo00} Bloser, P.~F., Grindlay, 
J.~E., Kaaret, P., Zhang, W., Smale, A.~P., Barret, D.\ 2000.\ RXTE Studies 
of Long-Term X-Ray Spectral Variations in 4U 1820-30.\ Astrophysical 
Journal 542, 1000-1015. 
 
\bibitem[Bradt et al.(1993)]{bra93} Bradt, H.~V., Rothschild, 
R.~E., Swank, J.~H.\ 1993.\ X-ray timing explorer mission.\ Astronomy and 
Astrophysics Supplement Series 97, 355-360. 

\bibitem[Kato(2001)]{kat01} Kato, S.\ 2001.\ Trapping of 
Non-Axisymmetric g-Mode Oscillations in Thin Relativistic Disks and kHz 
QPOs.\ Publications of the Astronomical Society of Japan 53, L37-L39. 
 
\bibitem[\protect\astroncite{van der Klis}{2005}]{kli05} van der Klis,
M.\ 2005, in Compact stellar X-ray  sources, Lewin \& van der Klis (eds),
Cambridge University Press, in press

\bibitem[van der Klis(2004)]{kli04} van der Klis, M.\ 2004.\ 
Neutron Star QPOs as Probes of Strong Gravity and Dense Matter.\ AIP 
Conf.~Proc.~714: X-ray Timing 2003: Rossi and Beyond 714, 371-378. 
 
\bibitem[Kluzniak et al.(1990)]{klu90} Kluzniak, W., 
Michelson, P., Wagoner, R.~V.\ 1990.\ Determining the properties of 
accretion-gap neutron stars.\ Astrophysical Journal 358, 538-544. 
 
\bibitem[Maccarone and Schnittman(2005)]{mac05} Maccarone, 
T.~J., Schnittman, J.~D.\ 2005.\ The bicoherence as a diagnostic for models 
of high-frequency quasi-periodic oscillations.\ Monthly Notices of the 
Royal Astronomical Society 357, 12-16. 
 
\bibitem[Miller et al.(1998)]{mil98} Miller, M.~C., Lamb, 
F.~K., Psaltis, D.\ 1998.\ Sonic-Point Model of Kilohertz Quasi-periodic 
Brightness Oscillations in Low-Mass X-Ray Binaries.\ Astrophysical Journal 
508, 791-830. 
 
\bibitem[Miller(2004)]{mil04} Miller, M.~C.\ 2004.\ 
Interpreting QPOs from Accreting Neutron Stars.\ AIP Conf.~Proc.~714: X-ray 
Timing 2003: Rossi and Beyond 714, 365-370. 

\bibitem[Muno(2004)]{mun04} Muno, M.~P.\ 2004.\ Millisecond 
Oscillations During Thermonuclear X-ray Bursts.\ AIP Conf.~Proc.~714: X-ray 
Timing 2003: Rossi and Beyond 714, 239-244. 
 
\bibitem[Miniutti et al.(2004)]{min04} Miniutti, G., Fabian, 
A.~C., Miller, J.~M.\ 2004.\ The relativistic Fe emission line in XTE 
J1650-500 with BeppoSAX: evidence for black hole spin and light-bending 
effects?.\ Monthly Notices of the Royal Astronomical Society 351, 466-472. 
 
\bibitem[Nowak(2004)]{now04} Nowak, M.~A.\ 2004.\ The Future 
of X-ray Spectroscopy of Galactic Black Hole Binaries.\ AIP 
Conf.~Proc.~714: X-ray Timing 2003: Rossi and Beyond 714, 89-96. 

\bibitem[Parmar et al.(2003)]{par03} Parmar, A.~N., Hasinger, G., Arnaud, M. et al., 2003.\ 
XEUS: the x-ray evolving universe spectroscopy mission.\ 
X-Ray and Gamma-Ray Telescopes and Instruments for Astronomy.~ Edited by 
Joachim E.~Truemper, Harvey D.~Tananbaum.~ Proceedings of the SPIE, Volume 
4851, pp.~304-313 (2003).\ 4851, 304-313. 
 
\bibitem[Poutanen(2004)]{pou04} Poutanen, J.\ 2004.\ The 
Physics of X-ray Emission from Accreting Millisecond Pulsars.\ AIP 
Conf.~Proc.~714: X-ray Timing 2003: Rossi and Beyond 714, 228-231. 
  
\bibitem[Psaltis(2004)]{psa04} Psaltis, D.\ 2004.\ 
Measurements of Black Hole Spins and Tests of Strong-Field General 
Relativity.\ AIP Conf.~Proc.~714: X-ray Timing 2003: Rossi and Beyond 714, 
29-35. 

\bibitem[Remillard(2004)]{rem04} Remillard, R.~A.\ 2004.\ 
X-ray QPOs from Black Hole Binary Systems.\ AIP Conf.~Proc.~714: X-ray 
Timing 2003: Rossi and Beyond 714, 13-20. 

\bibitem[Stella and Vietri(1999)]{ste99} Stella, L., Vietri, 
M.\ 1999.\ kHz Quasiperiodic Oscillations in Low-Mass X-Ray Binaries as 
Probes of General Relativity in the Strong-Field Regime.\ Physical Review 
Letters 82, 17-20. 
 
\bibitem[Swank(2004)]{swa04} Swank, J.\ 2004.\ Quasi-Periodic 
Oscillations from Low-mass X-Ray Binaries with Neutron Stars.\ AIP 
Conf.~Proc.~714: X-ray Timing 2003: Rossi and Beyond 714, 357-364. 
 
\bibitem[Strohmayer(2004)]{str04} Strohmayer, T.~E.\ 2004.\ 
Future Probes of the Neutron Star Equation of State Using X-ray Bursts.\ 
AIP Conf.~Proc.~714: X-ray Timing 2003: Rossi and Beyond 714, 245-252. 
 
\bibitem[Titarchuk and Wood(2004)]{tit04} Titarchuk, L., 
Wood, K.~S.\ 2004.\ Unified Model of Quasi-Periodic Oscillations.\ AIP 
Conf.~Proc.~714: X-ray Timing 2003: Rossi and Beyond 714, 383-386. 
 
\bibitem[Wijnands et al.(2003)]{wij03} Wijnands, R., van der 
Klis, M., Homan, J., Chakrabarty, D., Markwardt, C.~B., Morgan, E.~H.\ 
2003.\ Quasi-periodic X-ray brightness fluctuations in an accreting 
millisecond pulsar.\ Nature 424, 44-47. 
 
\bibitem[Zhang et al.(1998)]{zha98} Zhang, W., Smale, A.~P., 
Strohmayer, T.~E., Swank, J.~H.\ 1998.\ Correlation between Energy Spectral 
States and Fast Time Variability and Further Evidence for the Marginally 
Stable Orbit in 4U 1820-30.\ Astrophysical Journal 500, L171. 
 

\end{thebibliography}
\end{document}